\documentclass{appolb}
\usepackage{epsfig}

\newcommand{\revision}{}

\begin{document}
\pagestyle{plain}
\title{A singular perturbation approach to the steady-state 
1D Poisson-Nernst-Planck modeling}
\author{Ilona D. Kosi\'nska\address{
Institut f\"ur Physik, Universit\"at Augsburg\\
Universit\"atsstr. 1, 86159 Augsburg, Germany\\
M. Smoluchowski Institute of Physics, Jagiellonian University\\
Reymonta 4, 30-059 Krak\'ow, Poland}
\and
I. Goychuk, G. Schmid, P. H\"anggi\address{
Institut f\"ur Physik, Universit\"at Augsburg\\
Universit\"atsstr. 1, 86159 Augsburg, Germany
}
\and
M. Kostur\address{Institute of Physics, University of Silesia\\
Bankowa 14, 40-007 Katowice, Poland}
}
\maketitle

\begin{abstract}
The reduced 1D Poisson-Nernst-Planck (PNP) model of artificial nano\-pores in the presence 
of a permanent charge on the channel wall is studied. More specifically, we consider 
the limit where the channel length exceed 
much the Debye screening length and channel's charge is sufficiently small.
Ion transport is described 
by the nonequillibrium steady-state solution of 
the PNP system within a singular perturbation treatment.
The quantities, $1/\lambda$ -- the ratio of the Debye length to a 
characteristic length scale and $\epsilon$ -- the scaled intrinsic charge density, 
serve as the singular and the regular perturbation parameters, respectively.
The role of the boundary conditions is discussed. 
A comparison between numerics and the analytical results of the singular 
perturbation theory is presented. 
\end{abstract}


\section{Introduction}

In many physical situations an exact solution of the full problem could not be obtained, 
so that various asymptotic and perturbative techniques must be used. Moreover, in
many cases, even regular perturbation analysis is useless,
and singular perturbation methods must be applied.

An example of such a case constitutes a boundary-layer problem. It is a 1D differential-equation-boundary-value problem 
on the unit interval for
which the highest derivative of the differential equation is multiplied by a small parameter $\delta$:
\begin{equation}
\displaystyle \delta y''(x) + c(x) y'(x) + d(x) y(x)
= f(x),
\label{eq}
\end{equation}
with the boundary conditions
\begin{equation}
\begin{array}{cc}
y(0) = a, & y(1) = b.
\label{bc}
\end{array}
\end{equation}
This boundary-value problem is singular because in the limit $\delta\to 0$ one of the solutions abruptly
disappears and the limiting solution is not able to satisfy the two boundary conditions in (\ref{bc}).
This comes from the fact that by setting $\delta=0$, the order of the differential equation is reduced by 1. 
One of the method to solve the boundary-value problem (\ref{eq}) and (\ref{bc}) is to look for
solutions in the form of series expansion in powers of $\delta$. In order to construct an uniformly and globally valid solution,
the interval $0<x<1$ is decomposed into two kinds of regions, an outer region, in which the solution varies slowly as a function of $x$, and
an inner region or boundary-layer region, in which the solution varies rapidly as a function of $x$.
A boundary-layer region is a narrow region those thickness is typically of order $\delta$ or some power of $\delta$ \cite{Nayfeh}.
One of such boundary value problem is the 1D steady-state PNP system with non-vanishing
permanent surface charge on the walls of a nanopore.

Siwy and co-workers reported \cite{Siwy,Siwy2} that ion transport in nanopores
with asymmetric fixed charge distributions is characterized by such
interesting phenomena as ion current fluctuations, rectification, and pumping. 
For this reason, we analyse the boundary value problem of the
one-dimensional steady-state PNP system with
non-vanishing permanent surface charge.
Since the conical geometry is experimentally relevant,
we consider an uniformly charged conical pore. 

The flow of ions through the nanopore caused by an externally applied
electric field is analyzed by means of the Nernst-Planck equations
together with the Poisson equation, in a self-consistent manner. 
We are interested in ion
transport phenomena occuring in a very long channel i.e. in the limit
where the channel length exceed much the channel radii.  It justifies
the approximation of the channel as a one-dimensional object \cite{Kosinska}.  

The aim of the paper is to show an application of the singular perturbation 
theory to such one-dimensional
PNP systems where the following quantities - the ratio of the Debye length
$\xi_{\rm{D}}$ to the channel length $L$, denoted as $1/\lambda \sim
\xi_{\rm{D}}/L$, and the channel surface charge $\sigma$ - serve as the
perturbation parameters. The system can be viewed as a singularly
perturbed problem in $1/\lambda$ and a regularly perturbed in $\epsilon$
that denotes a dimensionless scaled surface charge density.
In the long channel limit, we analyzed the leading term
in $1/\lambda$, while $\epsilon$ is considered as the regular expansion
parameter. In our approach, we make use of above-described method to
construct an uniformly and globally valid solution by calculating separately
the outer and inner solutions and matching them then in a smooth manner.

\section{Ion Transport}

\subsection{One dimensional Poisson-Nernst -Planck (PNP) }

\begin{figure}[htb!]
\begin{center}
 \includegraphics[width=8.0cm]{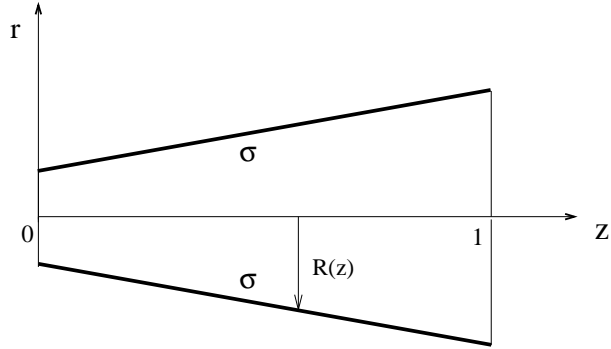}
\caption{Schema of the section of the conical channel. The channel's wall is uniformly charged with surface charge density $\sigma$. The local radius of the cone $R(z)$, is given by Eq.~\ref{radius}.}
\label{schema}
\end{center}
\end{figure}

In this paper we study ion transport through a long conical 
nanopore of the length $L = 1$ {\revision (scaling of the $z$-coordinate
by the channel length $L$, whereas the radial coordinate $r$ is scaled by 
the small opening radius $R(0)$)} 
with uniformly charged wall, cf.~Fig.~\ref{schema}.
The ion flow through the charged conical nanopore could be driven by 
the ion concentration gradients and by the electric field 
modeled together by means of the electrodiffusion equation. 
Diffusion in orthogonal direction of the pore is confined 
to a channel of variable area of $\pi R^2(z)$ (see Fig.~\ref{schema}). 
The electric field inside the pore is in turn governed by applied field and by the ion
concentrations through the Poisson equation. 
We are interested in the non-equillibrium steady-state
ion flux and electrical current which can persist either due to
applied voltage or due to a concentration gradient.

One can reduce the problem from 3D to 1D assuming instantaneous
equillibration in the transverse direction
\begin{equation}
\begin{array}{cc}
\bar{c}(z,t) \approx c(z,r,t) \mathcal{A}(z),~~ & \Phi(z,t) \approx \Phi(z,r,t),
\label{constant}
\end{array}
\end{equation}
where
\begin{equation} 
\begin{array}{cc}
\mathcal{A}(z) = \pi R^2(z),~~ &
R(z) = (1+\gamma z)
\label{radius} 
\end{array}
\end{equation}
with $\gamma = R(1) - 1$ denoting
{\revision a scaled slope of the cone's radius, 
with the scaling of the radial coordinate $r$ by $R(0)$}.

The 3D PNP equations for steady state ion currents $I_{\rm{K}^+}$ and 
$I_{\rm{Cl}^-}$
thus become in the 1D approximation (cf. Ref.~\cite{Kosinska} 
and Appendix A):

\begin{equation}
\begin{array}{cc}
\vspace{10pt}
\displaystyle - I_{\mathrm{K^+}} = 
\frac{d}{dz}\bar{c}_{\mathrm{K^+}}(z) + \bar{c}_{\mathrm{K^+}}(z)\frac{d\Phi(z)}{dz} - \bar{c}_{\mathrm{K^+}}(z)\frac{2}{R(z)}\frac{d R(z)}{dz},\\
\displaystyle  I_{\mathrm{Cl^-}} = 
\frac{d}{dz}\bar{c}_{\mathrm{Cl^-}}(z) - \bar{c}_{\mathrm{Cl^-}}(z)\frac{d\Phi(z)}{dz} - \bar{c}_{\mathrm{Cl^-}}(z)\frac{2}{R(z)}\frac{d R(z)}{dz},
\label{current}
\end{array}
\end{equation}
together with the reduced Poisson equation
\begin{equation}
\displaystyle \frac{d^2 \Phi(z)}{d z^{2}} + \frac{2}{R(z)}\frac{d R(z)}{dz}\frac{d\Phi(z)}{dz}
= -\lambda^2\frac{\epsilon}{R(z)} - \lambda^2 \frac{1}{\pi R^{2}(z)} \left( \bar{c}_{\mathrm{K^+}}(z) -  \bar{c}_{\mathrm{Cl^-}}(z)\right),
\label{poisson}
\end{equation}
where
\begin{equation}
\frac{1}{\lambda} = \displaystyle\sqrt{\frac{\epsilon_0\epsilon_w k_{\rm B} T R^2(0)}{\rm{e}^2 c_0 N_{\rm{A}} L^2}} = \sqrt{2c}\left(\frac{\xi_{\rm{D}}}{L}\right),
\label{lambda}
\end{equation}
is the scaled Debye screening length,
and the effective dimensionless surface charge density reads
\begin{equation}
\displaystyle \epsilon = \frac{2\sigma R(0)}{\rm{e} c_0 N_{\rm A}}.
\label{epsilon}
\end{equation}
In Eq.~(\ref{lambda}), $c$ denotes dimensionless concentration (see the definition given by Eq.~(\ref{dim_conc})),
$\xi_{\rm{D}} = \left[\epsilon_0\epsilon_w k_{\rm{B}} T/(2\rm{e}^2N_{\rm{A}} 10^3 c_{\rm{bulk}})\right]^{(1/2)}$ (in meters) 
is the Debye length in bulk (see Ref.~\cite{Kosinska,Jackson}), 
$\epsilon_0$ the dielectrical constant of vacuum,
$\epsilon_w\approx 80$ the relative dielectrical constant of
water, $k_{\rm{B}}$ the Boltzmann constant, $T$ the
temperature, $\rm{e}$ the elementary charge, $N_{\rm{A}}$ the Avogadro
number, $c_{\rm{bulk}}$ the bulk concentration of ions, and $c_0$ is an arbitrary reference 1D-concentration (in $1/6.023\times 10^{-14}$ moles/m). 
Note that the coordinate $z$ is also dimensionless in the above equations (measured in units of $L$).

\subsection{Rigorous Boundary Conditions}

The boundary conditions are
\begin{equation}
\begin{array}{cc}
\vspace{10pt}
\bar{c}_{\mathrm{K^+}}(0) = \pi c_{\mathrm{K^+,L}},& \bar{c}_{\mathrm{Cl^-}}(0) = \pi c_{\mathrm{Cl^-,L}},\\
\vspace{10pt}
\bar{c}_{\mathrm{K^+}}(1) = \pi (1+\gamma)^2 c_{\mathrm{K^+,R}},& \bar{c}_{\mathrm{Cl^-}}(1) = \pi (1+\gamma)^2c_{\mathrm{Cl^-,R}},\\
\Phi(0) = 0,& \Phi(1)=\Phi_{\mathrm{R}},
\label{boundary}
\end{array}
\end{equation}
wherein  
\begin{equation}
c_{i,\{\mathrm{L,R}\}} = 10^3 c_{\mathrm{bulk},\{\mathrm{L,R}\}}
R^2(0)/c_0,~~~\rm{i = K}^+,~\rm{Cl}^- 
\label{dim_conc}
\end{equation}
and $c_{\mathrm{bulk}, \{\mathrm{L, R}\}}$ denote the bulk
concentrations of ions on the left and right sides (in moles), respectively.
The electro-neutrality condition yields: $c_{\mathrm{K^+,L}}=c_{\mathrm{Cl^-,L}}=c_{\mathrm{L}}$, and $c_{\mathrm{K^+,R}}=c_{\mathrm{Cl^-,R}}=c_{\mathrm{R}}$.
Furthermore, the difference of dimensionless potentials across the
nanopore is related to the applied voltage $U$ (in units of  Volts)
by $\Phi(0)-\Phi(1) = \mathrm{e} U / (k_{\mathrm{B}} T) $, yielding $U = - k_{\rm{B}}T\Phi(1)/\rm{e}$.

\section{Singular perturbation study}

\subsection{Perturbation parameters}

The coresponding system contains two small parameters.
The first one $1/\lambda$ is related to the ratio of the Debye length to a 
characteristic length scale.
Assuming that $\lambda\to \infty$ we search for an aproximate solution of equations (\ref{current}) and (\ref{poisson}) in the form
\begin{equation}
\begin{array}{c}
\vspace{10pt}
\displaystyle\Phi (z) = \Phi^{(0)}(z) + \frac{1}{\lambda}\Phi^{(1)}(z) + \dots,\\
\vspace{10pt}
\displaystyle \bar{c}_{\mathrm{K^+}}(z) = c_{\mathrm{K^+}}^{(0)}(z) + \frac{1}{\lambda} c_{\mathrm{K^+}}^{(1)}(z) + \dots,\\
\displaystyle \bar{c}_{\mathrm{Cl^-}}(z) = c_{\mathrm{Cl^-}}^{(0)}(z) + \frac{1}{\lambda} c_{\mathrm{Cl^-}}^{(1)}(z) + \dots.
\end{array}
\end{equation}
The second small parameter corresponds to the re-scaled intrinsic charge density $\epsilon$ (see in Eq.~\ref{epsilon}).
Furthermore, we represent $\Phi^{(i)}(z), c_{\mathrm{K^+}}^{(i)}(z), c_{\mathrm{Cl^-}}^{(i)}(z)$ as a series in $\epsilon$
\begin{equation}
\begin{array}{c}
\vspace{10pt}
\displaystyle\Phi^{(i)}(z) = \Phi^{(i)}_{(0)}(z) - \epsilon\Phi^{(i)}_{(1)}(z) +\dots,\\
\vspace{10pt}
\displaystyle \bar{c}_{\mathrm{K^+}}^{(i)}(z) = c_{\mathrm{K^+},~{(0)}}^{(i)}(z) - \epsilon c_{\mathrm{K^+},~{(1)}}^{(i)}(z) + \dots,\\
\displaystyle \bar{c}_{\mathrm{Cl^-}}^{(i)}(z) = c_{\mathrm{Cl^-}, ~{(0)}}^{(i)}(z) - \epsilon c_{\mathrm{Cl^-},~{(1)}}^{(i)}(z) + \dots.
\end{array}
\end{equation}
where $i = 0,1,\dots$.
Finally, a singularly perturbated boundary --
value problem in two independent perturbation parameters $1/\lambda$
and $\epsilon$ is obtained. We search for the solution letting
$1/\lambda\to 0$, while $\epsilon$ is considered as the expansion
parameter. Note that the perturbation problem is singular in $1/\lambda$ and regular in $\epsilon$.

\subsection{The outer expansion.}

These series are obtained by holding $z$ fixed and letting $1/\lambda\to 0$.
In zeroth order in $1/\lambda$, the problem reduces to
\begin{equation}
\left( c^{(0)}_{\rm{K}^+}(z) - c^{(0)}_{\rm{Cl}^-}(z) \right)/\pi R^{2}(z) = -\frac{\epsilon}{R(z)},
\label{conc_dwa}
\end{equation}
reflecting the charge neutrality in the interior, and
\begin{equation}
\begin{array}{c}
\vspace{20pt}
\displaystyle -I^{(0)}_{\rm{K}^+} =  \frac{d}{dz} c^{(0)}_{\rm{K}^+}(z) -2\frac{R'(z)}{R(z)} c^{(0)}_{\rm{K}^+}(z) +
 \frac{d\Phi^{(0)}(z)}{dz} c^{(0)}_{\rm{K}^+}(z),\\
\displaystyle I^{(0)}_{\rm{Cl}^-} =  \frac{d}{dz} c^{(0)}_{\rm{Cl}^-}(z) - 2\frac{R'(z)}{R(z)}c^{(0)}_{\rm{Cl}^-}(z) - \frac{d\Phi^{(0)}(z)}{dz} c^{(0)}_{\rm{Cl}^-}(z).
\label{currents}
\end{array}
\end{equation}

Adding and substracting the current formulas, we obtain
\begin{equation}
\begin{array}{c}
\vspace{10pt}
\displaystyle -I^{(0)}_{\rm{K}^+} + I^{(0)}_{\rm{Cl}^-} = 
\frac{d}{dz} \left( c^{(0)}_{\rm{K}^+}(z) + c^{(0)}_{\rm{Cl}^-}(z)\right)
\\
\vspace{10pt}\displaystyle -2\frac{R'(z)}{R(z)}\left( c^{(0)}_{\rm{K}^+}(z)+c^{(0)}_{\rm{Cl}^-}(z) \right)
- \frac{d\Phi^{(0)}}{dz} \epsilon \pi R(z),\\
\displaystyle -\left(I^{(0)}_{\rm{K}^+}+I^{(0)}_{\rm{Cl}^-}\right) = -\epsilon \pi R'(z) + \frac{d\Phi^{(0)}}{dz} \left(c^{(0)}_{\rm{K}^+}(z) + c^{(0)}_{\rm{Cl}^-}(z)\right),
\end{array}
\end{equation}
where $-I^{(0)}_{\rm{K}^+}+I^{(0)}_{\rm{Cl}^-} = J^{(0)} = \rm{const}$, $I^{(0)}_{\rm{K}^+}+I^{(0)}_{\rm{Cl}^-} = I^{(0)} = \rm{const}$.
The both sorts of ions have equal diffusion coefficients
(see Ref.~\cite{book}), i.e. $D_{\mathrm{\rm{K}^+}}=D_{\mathrm{\rm{Cl}^-}} = D$, thus $J^{(0)}$ yields an approximation 
to the negative of the total mass
current, whereas $I^{(0)}$ approximates the total electric current (see also Ref.~\cite{Barcilon}),
\begin{equation}
\begin{array}{c}
\vspace{20pt}
\displaystyle \pi\epsilon R(z)\frac{d\Phi^{(0)}(z)}{dz} =  - J^{(0)} +  \frac{d}{dz} \left( c^{(0)}_{\rm{K}^+}(z) + c^{(0)}_{\rm{Cl}^-}(z)\right)\\
\vspace{10pt}\displaystyle  - 2\frac{R'(z)}{R(z)}\left( c^{(0)}_{\rm{K}^+}(z)+c^{(0)}_{\rm{Cl}^-}(z) \right),\\
\displaystyle \left(c^{(0)}_{\rm{K}^+}(z) + c^{(0)}_{\rm{Cl}^-}(z)\right)\frac{d\Phi^{(0)}(z)}{dz} = - I^{(0)} +  \pi\epsilon  R'(z).
\end{array}
\end{equation}
Then one gets
\begin{equation}
\begin{array}{c}
\vspace{10pt}
\displaystyle  \left(c^{(0)}_{\rm{K}^+}(z) + c^{(0)}_{\rm{Cl}^-}(z)\right)\bigg( J^{(0)} -  \frac{d}{dz} \left( c^{(0)}_{\rm{K}^+}(z) + c^{(0)}_{\rm{Cl}^-}(z)\right) \\
\vspace{10pt}\displaystyle
+ 2\frac{R'(z)}{R(z)}\left( c^{(0)}_{\rm{K}^+}(z)+c^{(0)}_{\rm{Cl}^-}(z) \right)\bigg) = 
\displaystyle \pi\epsilon R(z) \left(
  I^{(0)} - \pi\epsilon R'(z)
\right).
\end{array}
\end{equation}
One cannot find exact solution to this differential equation. It is solved perturbatively in $\epsilon$
with the solution approximated as
\begin{equation}
\begin{array}{c}
\vspace{10pt}
\displaystyle \left( c^{(0)}_{\rm{K}^+}(z) + c^{(0)}_{\rm{Cl}^-}(z)\right) = c^{(0)}_{\Sigma}(z) = c^{(0)}_{\Sigma, (0)}(z) - \epsilon  c^{(0)}_{\Sigma, (1)}(z) + \dots\\
\Phi^{(0)}(z) = \Phi^{(0)}_{(0)}(z) -\epsilon \Phi^{(0)}_{(1)}(z) + \dots,
\label{outer_sol}
\end{array}
\end{equation}
where the upper index denotes the order of expansion in $1/\lambda$, and the lower one in $\epsilon$.

\subsection{The boundary layer solutions.}

Let $\delta(1/\lambda)$ be the width of the boundary layer which is a function of $1/\lambda$. The
problem is rescaled near $z = 0$ by setting $\zeta =
z/\delta(1/\lambda) $. It has been found that $\delta(1/\lambda) =
1/\lambda$ and for $z$ in the boundary layer, $\zeta = O(1)$. Thus, we
introduce the stretched coordinate
\begin{equation}
\zeta = \lambda z,
\end{equation}
where $\zeta\in [0, +\infty)$ and express the concentrations and the electric potential in terms of this coordinate as
\begin{equation}
\begin{array}{c}
\vspace{10pt}
\displaystyle c_{\rm{K}^+}(\zeta/\lambda;1/\lambda) = 
\displaystyle p(\zeta;1/\lambda),\\
\vspace{10pt}
\displaystyle c_{\rm{Cl}^-}(\zeta/\lambda;1/\lambda) = 
\displaystyle n(\zeta;1/\lambda),\\
\displaystyle \Phi(\zeta/\lambda;1/\lambda) = 
\displaystyle \phi(\zeta;1/\lambda).
\end{array}
\end{equation}
In the right boundary layer, i.e. near $z=1$ we set the another stretched variable
\begin{equation}
\displaystyle \chi = \lambda (z - 1),
\end{equation}
where $\chi\in (-\infty,0]$
and define analogously
\begin{equation}
\begin{array}{c}
\vspace{10pt}
\displaystyle c_{\rm{K}^+}(\chi/\lambda+1;1/\lambda) = 
\displaystyle \tilde{p}(\chi;1/\lambda),\\
\vspace{10pt}
\displaystyle c_{\rm{Cl}^-}(\chi/\lambda +1;1/\lambda) = 
\displaystyle \tilde{n}(\chi;1/\lambda),\\
\displaystyle \Phi(\chi/\lambda +1;1/\lambda) = 
\displaystyle \tilde{\phi}(\chi;1/\lambda).
\end{array}
\end{equation}
Each of these functions is written as an asymptotic series, namely,
\begin{equation}
\begin{array}{cc}
\vspace{10pt}
\displaystyle p(\zeta; 1/\lambda) = p^{(0)}(\zeta) + \frac{1}{\lambda} p^{(1)}(\zeta)+\dots,& ~~\displaystyle \tilde{p}(\chi;1/\lambda) = \tilde{p}^{(0)}(\chi) + \frac{1}{\lambda} \tilde{p}^{(1)}(\chi)+\dots,\\
\vspace{10pt}
\displaystyle n(\zeta; 1/\lambda) = n^{(0)}(\zeta) + \frac{1}{\lambda} n^{(1)}(\zeta)+\dots,& ~~ \displaystyle \tilde{n}(\chi;1/\lambda) = \tilde{n}^{(0)}(\chi) + \frac{1}{\lambda} \tilde{n}^{(1)}(\chi)+\dots,\\
\displaystyle 
\phi (\zeta; 1/\lambda) = \phi^{(0)}(\zeta) + \frac{1}{\lambda}\phi^{(1)}(\zeta)+\dots,& ~~ \displaystyle \tilde{\phi}(\chi;1/\lambda) = \tilde{\phi}^{(0)}(\chi) + \frac{1}{\lambda}\tilde{\phi}^{(1)}(\chi)+\dots,
\end{array}
 \end{equation}
and 
\begin{equation}
\begin{array}{cc}
\vspace{10pt}
\displaystyle   R(\zeta) \to  1, & \quad R(\chi) \to 1+\gamma.
\end{array}
\end{equation}
The upper indixes  correspond to the order of the expansion in $1/\lambda$.
The limit expansions is obtained by holding $\zeta$ and $\chi$ fixed and letting $1/\lambda\to 0$.

In turn, each of the $p^{(0)}(\zeta), n^{(0)}(\zeta), \phi^{(0)}(\zeta)$ functions is written as an asymptotic series in $\epsilon$, namely,
\begin{equation}
\begin{array}{cc}
\vspace{10pt}
\displaystyle p^{(0)}(\zeta) = p^{(0)}_{(0)}(\zeta) - \epsilon p^{(0)}_{(1)}(\zeta)+\dots,& \quad \displaystyle \tilde{p}^{(0)}(\chi) = \tilde{p}^{(0)}_{(0)}(\chi) - \epsilon \tilde{p}^{(0)}_{(1)}(\chi)+\dots,\\
\vspace{10pt}
\displaystyle n^{(0)}(\zeta) = n^{(0)}_{(0)}(\zeta) - \epsilon n^{(0)}_{(1)}(\zeta)+\dots,& \quad \displaystyle \tilde{n}^{(0)}(\chi) = \tilde{n}^{(0)}_{(0)}(\chi) - \epsilon \tilde{n}^{(0)}_{(1)}(\chi)+\dots,\\
\displaystyle 
\phi^{(0)}(\zeta) = \phi^{(0)}_{(0)}(\zeta) - \epsilon\phi^{(0)}_{(1)}(\zeta)+\dots,& \quad \displaystyle \tilde{\phi}^{(0)}(\chi) = \tilde{\phi}^{(0)}_{(0)}(\chi) - \epsilon\tilde{\phi}^{(0)}_{(1)}(\chi)+\dots.
\label{lb_solutions}
\end{array}
 \end{equation}
The lower indixes refer to the order of the series expansion in $\epsilon$.
In the leading order functions in $1/\lambda$ the transport equations read
\begin{equation}
\begin{array}{cc}
\vspace{20pt}
\displaystyle 0 = \frac{d}{d\zeta} p^{(0)}(\zeta) +
\frac{d\phi^{(0)}(\zeta)}{d\zeta}p^{(0)}(\zeta),& \quad \displaystyle 0 = \frac{d}{d\chi} \tilde{p}^{(0)}(\chi) +
\frac{d\tilde{\phi}^{(0)}(\chi)}{d\chi}\tilde{p}^{(0)}(\chi),\\
\displaystyle 0 =  \frac{d}{d\zeta} n^{(0)}(\zeta) - 
\frac{d\phi^{(0)}(\zeta)}{d\zeta}n^{(0)}(\zeta), & \quad \displaystyle 0 =  \frac{d}{d\chi} \tilde{n}^{(0)}(\chi) - 
\frac{d\tilde{\phi}^{(0)}(\chi)}{d\chi}\tilde{n}^{(0)}(\chi),
\label{NP}
\end{array}
\end{equation}
together with the Poisson equation
\begin{equation}
\begin{array}{c}
\vspace{10pt}
\displaystyle \frac{1}{\pi}\left(p^{(0)}(\zeta) - n^{(0)}(\zeta)\right) + \epsilon = -\frac{d^2\phi^{(0)}(\zeta)}{d\zeta^2},\\
\displaystyle \frac{1}{\pi (1+\gamma)^2}\left(\tilde{p}^{(0)}(\chi) - \tilde{n}^{(0)}(\chi)\right) + \frac{\epsilon}{1+\gamma} = -\frac{d^2\tilde{\phi}^{(0)}(\chi)}{d\chi^2},
\label{poisson_boundary}
\end{array}
\end{equation}
and obeys the following boundary conditions
\begin{equation}
\begin{array}{cccc}
\vspace{10 pt}
 p^{(0)}_{(0)}(0) = \pi c_{\mathrm{L}},& p^{(0)}_{(i)}(0) =0,& \quad \tilde{p}^{(0)}_{(0)}(0) = \pi (1+\gamma)^2 c_{\mathrm{R}}, &
\tilde{p}^{(0)}_{(i)}(0) = 0, \\
\vspace{10pt}
 n^{(0)}_{(0)}(0) = \pi c_{\mathrm{L}},& n^{(0)}_{(i)}(0) =0,& \quad \tilde{n}^{(0)}_{(0)}(0) = \pi (1+\gamma)^2 c_{\mathrm{R}}, &
\tilde{n}^{(0)}_{(i)}(0) = 0, \\
 \phi^{(0)}_{(0)}(0) = 0,&\phi^{(0)}_{(i)}(0) = 0,& \quad
 \tilde{\phi}^{(0)}_{(0)}(0) = \Phi_{\mathrm{R}}, & \tilde{\phi}^{(0)}_{(i)}(0) = 0,
\end{array}
\end{equation}
with $i = 1, 2, \dots$. As a next step, we integrate the left boundary layer of the Nernst-Planck equations (\ref{NP})
\begin{equation}
\begin{array}{cc}
\vspace{10pt}
\displaystyle p^{(0)}(\zeta) = p^{(0)}_{(0)}(0) e^{-\phi^{(0)}_{(0)}(\zeta)} \left\{ 1 + \epsilon
\phi^{(0)}_{(1)}(\zeta) + \epsilon^2
\left(-\phi^{(0)}_{(2)}(\zeta)+\frac{1}{2}\phi^{(0) 2}_{(1)}(\zeta)\right) \right\} + O(\epsilon^3),\\
\displaystyle n^{(0)}(\zeta) = n^{(0)}_{(0)}(0) e^{\phi^{(0)}_{(0)}(\zeta)} \left\{ 1 - \epsilon \phi^{(0)}_{(1)}(\zeta) + 
\epsilon^2 \left(\phi^{(0)}_{(2)}(\zeta) + \frac{1}{2}\phi^{(0) 2}_{(1)}(\zeta)\right)\right\} + O(\epsilon^3).
\label{solNP}
\end{array}
\end{equation}
After the substitution $\{p,n,\phi\}\to\{\tilde{p}, \tilde{n},
\tilde{\phi}\}$ in the equations (\ref{solNP}), we obtain the right boundary layer solutions.
Then, we put these expressions for the concentrations into the Poisson 
equation (\ref{poisson_boundary}).

\subsection{Matching procedure.} 
Now a matching of these two representations is needed. So we choose
constants to make left (right) boundary layer solutions and outer
expansion, respectively, coincide for each order of $\epsilon$ (as $1/\lambda\to 0$) in some intermediate zone
between the left (right) boundary layer and the outer region,
respectively. Furthemore, in the left intermediate zone an
intermediate variable $z_\alpha$ is introduced, where
\begin{equation}
z_\alpha = \left\{
\begin{array}{c}
\vspace{10pt}
 (1/\lambda)^{-\alpha}z,\\
 (1/\lambda)^{1-\alpha}\zeta,
\end{array}
\right.
\end{equation}
and $0<\alpha<1$ such as
\begin{equation}
\begin{array}{c}
\vspace{10pt}
\displaystyle \lim \frac{\zeta}{z_\alpha} \to \infty ~~\rm{when}~~\frac{1}{\lambda}\to 0,\\
\displaystyle \lim \frac{z}{z_\alpha} \to 0 ~~\rm{when}~~\frac{1}{\lambda}\to 0.
\end{array}
\end{equation}
The above-described procedure is also applied to find an intermediate variable ($z_\beta - 1 = (1/\lambda)^{1-\beta}\chi$) in the right intermediate zone. 
The ensuing conditions are ($ z_\alpha, z_\beta ~\rm{fixed ~ and}~1/\lambda\to 0$)
\begin{equation}
\begin{array}{cc}
\vspace{10pt}
\displaystyle c^{(0)}_{\rm{K}^+}(0) = p^{(0)}(\infty), & \quad c^{(0)}_{\rm{K}^+}(1) = \tilde{p}^{(0)}(-\infty),\\
\vspace{10pt}
c^{(0)}_{\rm{Cl}^-}(0) =  n^{(0)}(\infty), & \quad c^{(0)}_{\rm{Cl}^-}(1) =  \tilde{n}^{(0)}(-\infty),\\
\Phi^{(0)}(0) = \phi^{(0)}(\infty), & \quad \Phi^{(0)}(1) = \tilde{\phi}^{(0)}(-\infty),
\label{match}
\end{array}
\end{equation}
and
\begin{equation}
\begin{array}{cc}
\displaystyle \frac{d\phi^{(0)}}{d\zeta}(\infty) = 0, & \quad \displaystyle \frac{d\tilde{\phi}^{(0)}}{d\chi}(-\infty) = 0.
\end{array}
\end{equation}
Adding and substracting the formulas (\ref{conc_dwa}) and (\ref{con_0}), (\ref{conc_dwa}) and (\ref{con_1}), respectively, we obtain
\begin{equation}
\displaystyle c^{(0)}_{(i), \{\mathrm{K^+, Cl^-}\}}(0) = \frac{1}{2} c^{(0)}_{(i), \Sigma}(0) \pm 
\left\{
\begin{array}{cc}
\vspace{10pt}
\displaystyle \frac{\pi}{2} & \rm{for, i = 1}\\
\displaystyle 0 & \rm{for, i = 0,2,3,\dots}
\end{array}
\right.
\label{conditions}
\end{equation}
where the upper sign refers to K$^+$ ions and the lower one to Cl$^-$.
To find the values of $\phi^{(0)}(\infty)$, we make use of 
\begin{equation}
\displaystyle \frac{1}{\pi}\left(p^{(0)}(\infty) - n^{(0)}(\infty) \right) + \epsilon = 0.
\end{equation}
Taking into account the matching conditions (\ref{match}) together
with (\ref{conditions}) we get conditions for unknown
constants $J^{(0)}_{(i)},I^{(0)}_{(i)},C^{(0)}_{(i)},E^{(0)}_{(i)}$, where
$i=0,1,2,\dots$

The above procedure allows us to find matching conditions between the outer expansion and the right boundary layer as well.

\subsection{The Poisson equation in the boundary layers.}

Now we consider the Poisson
equation (\ref{poisson_boundary}) within the boundary layer, and integrate them following the Ref.
\cite{Barcilon}.
We are able to find analytical solutions of (\ref{poisson_boundary})
up to the second order in $\epsilon$. The solutions of the Poisson equation
(\ref{poisson_boundary}) depend on the boundary conditions. 
For $c_{\rm{K}^+,L}=c_{\rm{Cl}^-,L}$,
$\phi^{(0)}_{(0)}(\zeta)=0$ (see also Ref.~\cite{Barcilon}),
$\displaystyle\phi^{(0)}_{(1)}(\zeta) =\frac{1}{2c_L}\left( e^{-\sqrt{2c_L}\zeta} - 1  \right)$, and $\phi^{(0)}_{(2)}(\zeta)=0$.
 
Similar results hold for the solutions in the right boundary layer. If $c_{\rm{K}^+,R}=c_{\rm{Cl}^-,R}$, then
$\tilde{\phi}^{(0)}_{(0)}(\chi)=\Phi_R$
(as in Ref.~\cite{Barcilon}), \\ $\displaystyle\tilde{\phi}^{(0)}_{(1)}(\chi)=
\frac{1}{2c_R(1+\gamma)}\left( e^{\sqrt{2c_L}\chi} - 1 \right)$, and
$\tilde{\phi}^{(0)}_{(2)}(\chi)=0$.

\subsection{Uniformly valid solutions.}

To obtain an approximation that is valid uniformly on $[0, 1]$, we add
the boundary and outer appoximations and subtract their common limit
in the intermediate zone. Thus, we find:
\begin{eqnarray}
\vspace{10pt}
\Phi(z) = 
\Phi^{(0)}_{(0)}(z) &-&
\epsilon 
\bigg\{
\phi^{(0)}_{(1)}(\lambda z) +
\Phi^{(0)}_{(1)}(z)+ 
\tilde{\phi}^{(0)}_{(1)}(\lambda (1-z)) \nonumber \\
&+&\displaystyle\frac{1}{2c_{\mathrm{L}}}+
\displaystyle
\frac{1}{2c_{\mathrm{R}}(1+\gamma)}\bigg\} + O(\epsilon^2),
\label{pot_uni}
\end{eqnarray}
{}
\begin{eqnarray}
\vspace{10pt}
\displaystyle \bar{c}_{\{\mathrm{K^+, Cl^-}\}}(z)& = &\frac{1}{2}
c^{(0)}_{\Sigma, (0)}(z)
- \frac{1}{2} \epsilon\bigg\{ \pm\pi R(z) 
+ c^{(0)}_{\Sigma, (1)}(z) \mp
2\pi c_{\mathrm{L}}\phi^{(0)}_{(1)}(\lambda z) \nonumber \\ 
&\mp& 2\pi c_{\mathrm{R}} (1+\gamma)^2
\tilde{\phi}^{(0)}_{(1)}(\lambda (1-z)) \mp \pi(2+\gamma) \bigg\}
 + O(\epsilon^2),
\label{con_pos_uni}
\end{eqnarray}
where the upper sign refers to K$^+$ ions and the lower one to Cl$^-$.

\section{The Donnan boundary conditions}

Approximatively, one can impose the so-called Donnan equilibrium
boundary conditions at the channel ends \cite{Cervera}, reading:
\begin{eqnarray}
\vspace{10pt}
\displaystyle
\bar{c}^{\mathrm{Don}}_i(z_ {\{\mathrm{ L, R}\}}) = 
\frac{10^3 \pi R(z_{\{\mathrm{L, R} \}})^2}{2 c_0} \Big( -\nu_i X_{\{ \mathrm{L, R}\}} 
 +  \sqrt{X^2_{\{\mathrm{L, R}\}}
+ 4 c^2_{\mathrm{bulk}, \{\mathrm{L, R}\}} } \Big),
\label{Don_1}
\end{eqnarray}
\begin{eqnarray}
\displaystyle\Phi^{\mathrm{Don}}(z_ {\{\mathrm{ L, R}\}})  = 
\Phi_{\{\mathrm{L, R}\}} - \frac{1}{\nu_i}\ln
\frac{\bar{c}^{\mathrm{Don}}_i(z_ {\{ \mathrm{L, R}\}}) c_0}{10^3 \pi R(z_{\{\mathrm{L, R} \}})^2
c_{\mathrm{bulk}}(z_ {\{ \mathrm{L, R}\}})},
\label{Don_2}
\end{eqnarray}
where $\displaystyle X_{\{\mathrm{L, R}\}} = \displaystyle\frac{2\sigma}{F}
\times\frac{1}{R(z_{\{\mathrm{L, R}\}})}$ (in M) and i = K$^+$, Cl$^-$. Here,  $R(z_{\{ \mathrm{L,
R}\}})$ denotes the nanopore radii at the left and right ends of the
pore, respectively.

Let us expand the Donnan equilibrium conditions 
in terms of power of $\epsilon$ 
and afterwards, using the definition of $c_{i,{\{\mathrm{L,R}\}}}$, i = K$^+$, Cl$^-$ 
(see Sec. Rigorous Boundary Conditions), 
we find our dimensionless functions at both channel ends
\begin{equation}
\begin{array}{c}
\vspace{10pt}
\displaystyle \bar{c}^{\rm{Don}}_i(z_{\{\mathrm{L, R}\}}) = \pi R(z_{\{\mathrm{L, R} \}})^2 c_{i} \mp \frac{\pi}{2}R(z_{\{\mathrm{L, R}\}})\epsilon \\
\vspace{10pt}\displaystyle + \frac{\pi}{8 c_{i}}\epsilon^2 - \frac{\pi}{128 c^3_{i} R(z_{\{\mathrm{L, R}\}})^2}\epsilon^4 + \dots,\\
\displaystyle \Phi^{\rm{Don}}(z_{\{\mathrm{L, R}\}}) = \Phi_{\{\mathrm{L, R}\}} + \frac{1}{2c_{i}R(z_{\{\mathrm{L, R}\}})}\epsilon \\
\vspace{10pt}\displaystyle - \frac{1}{48c^3_{i}R(z_{\{\mathrm{L, R}\}})^3}\epsilon^3 + \frac{3}{1280 c^5_{i} R(z_{\{\mathrm{L, R}\}})^5}\epsilon^5 +\dots,\\
\label{Donnan}
\end{array}
\end{equation}
where the upper sign in $\bar{c}^{\rm{Don}}_i(z_{\{\mathrm{L, R}\}})$ refers to
potassium ions and the lower one to chloride, $i=\rm{K}^+,\rm{Cl}^-$, $z_\mathrm{L}=1$, and 
$z_{\mathrm{R}}=(1+\gamma)$. Note that the Donnan equilibrium conditions (Eq.
(20)-(21) in Ref.~\cite{Cervera}) have been obtained under the assumption
of local electroneutrality at the channel borders.

{\bf Theorem 1}

\emph{Let $1/\lambda\to 0$. If $\zeta\to + \infty$, then the left
  boundary layer solutions
  $p^{(0)}(\zeta),n^{(0)}(\zeta),\phi^{(0)}(\zeta)$ given by an
  asymptotic series (\ref{lb_solutions}) tend to the Donnan asymptotic
  series (\ref{Donnan}) in $z_L$.}

One can formulate analogue of above theorem for right boundary layer solutions.

Proof: We first observe that the solution of the following transport equation
\begin{eqnarray}
\displaystyle - I^{(0)}_p & = &\frac{d}{d\zeta} p^{(0)}(\zeta) +
\frac{d\phi^{(0)}}{d\zeta}p^{(0)}(\zeta),\quad\quad \zeta\in [0,+\infty),\\
\displaystyle  I^{(0)}_n & = &\frac{d}{d\zeta} n^{(0)}(\zeta) -
\frac{d\phi^{(0)}}{d\zeta}n^{(0)}(\zeta),
\end{eqnarray}
when $\displaystyle I^{(0)}_{\{p, n\}} \to 0$ (the equilibrium condition) tends to the Boltzmann distribution:
\begin{eqnarray}
p^{(0)}(\zeta) & \to & p^{(0)}(0) e^{-(\phi^{(0)}(\zeta) - \phi^{(0)}(0))},\\
n^{(0)}(\zeta) & \to & n^{(0)}(0) e^{(\phi^{(0)}(\zeta) - \phi^{(0)}(0))}.
\end{eqnarray}
It gives us the Donnan equilibrium condition for a given potential drop. 
On the other hand, it is known from the matching procedure that
$p^{(0)}(\zeta)$, $n^{(0)}(\zeta)$, $\phi^{(0)}(\zeta)$ when $\zeta\to
+\infty$ must tend to the outer expansions $\bar{c}^{(0)}_{\rm{K}^+}(0)$, 
$\bar{c}^{(0)}_{\rm{Cl}^-}(0)$, $\Phi^{(0)}(0)$. This means that for $\zeta\to
+\infty$ the condition of local electroneutrality is fulfilled. 
And then we end up with series given by Eqs.~(\ref{Donnan}).

In turn, the outer solutions
$\bar{c}^{(0)}_{\rm{K}^+}(z)$, $\bar{c}^{(0)}_{\rm{Cl}^-}(z)$, $\Phi^{(0)}(z)$ (given by (\ref{con_pos_uni}) and (\ref{pot_uni})) 
tend to the Donnan asymptotic series (\ref{Donnan}) for $z=z_{\{\mathrm{L,R}\}}$, respectively.
Thus, on the basis of {\bf Theorem 1} and the matching
procedure, one can formulate

{\bf Theorem 2}

\emph{If $1/\lambda\to 0$, then the outer expansions
  $\bar{c}^{(0)}_{\rm{K}^+}(z), \bar{c}^{(0)}_{\rm{Cl}^-}(z), \Phi^{(0)}(z)$ approximate the
  solutions of the 1D-PNP system (Eq. (\ref{poisson}) and
  (\ref{current})) with the Donnan boundary conditions satisfying the
  local electroneutrality at the boundaries.}

Proof: 
 
The boundary layer solutions calculated with the Donnan boundary conditions (\ref{Donnan})
(after inversion to normal coordinate $z$) are restricted to the points
at the boundaries approximated by these series (\ref{Donnan}).

The above-presented theorem was already demonstrated in
Ref.~\cite{Gillespie}, in which a singular
perturbation expansion was also used to derive the Donnan potential. We
recall this finding, however, in a different manner. 
One can conclude that in the limit where the channel length exceeds much
the Debye screening length, the Donnan jumps at the channel borders can
be safely used instead of the rigorous treatment of the   the boundary
layers of finite width. 

Moreover, on the basis of the matching conditions given by (\ref{match})
and {\bf Theorem 1}, we notice that unknown constants (present in the outer approximation), 
in particular the total flux $J^{(0)}$ and the electric current $I^{(0)}$,
in both discussed cases of boundary conditions constitute the same series in $\epsilon$.
 
\section{Results}

\subsection{Perturbation vs. numerics}

We shall present a comparison of the analytical results calculated by
the above-described singular perturbation method with numerical
solutions. The 1D PNP system (equations (\ref{current})-(\ref{poisson})),
with boundary conditions (equations in (\ref{boundary})), is 
integrated numerically by making use of a collocating method with adaptive meshing \cite{NAG}. 
We choose the following set of parameters:
$R(0)=3$ nm, $R(1)=6.616$ nm, $L=200$ nm, $\gamma = 1.2055$, the room
temperature ($T=298$ K), the relative dielectric constant of water
$\epsilon_w=80$, and the diffusion coefficients of the ions
$D_{\rm{K}^+}=D_{\rm{Cl}^-} = 2\times 10^{9}$ nm$^2$/s \cite{book}. For such a
channel, the value of the first perturbation parameter $1/\lambda$ is
about $10^{-2}$. The square of this parameter is sufficiently small.
However, this is not enough to provide a good approximation to the PNP
system with a varying value of the surface charge density which is not 
that small in the reality. 
To test the
limits of regular perturbation method with respect to the
second expansion parameter $\epsilon$, we consider the two different
cases (a) $\sigma=-0.02$ e/nm$^2$, and (b) $\sigma=-0.1$ e/nm$^2$. The
first one corresponding to $\epsilon = 0.12$ is well within the
perturbative treatment. For the second one with $\epsilon = 0.6$ the
perturbative treatment is expected to fail, but it might work
occasionally.

\subsubsection{Uniformly valid approximation}

\begin{figure}[htb!]
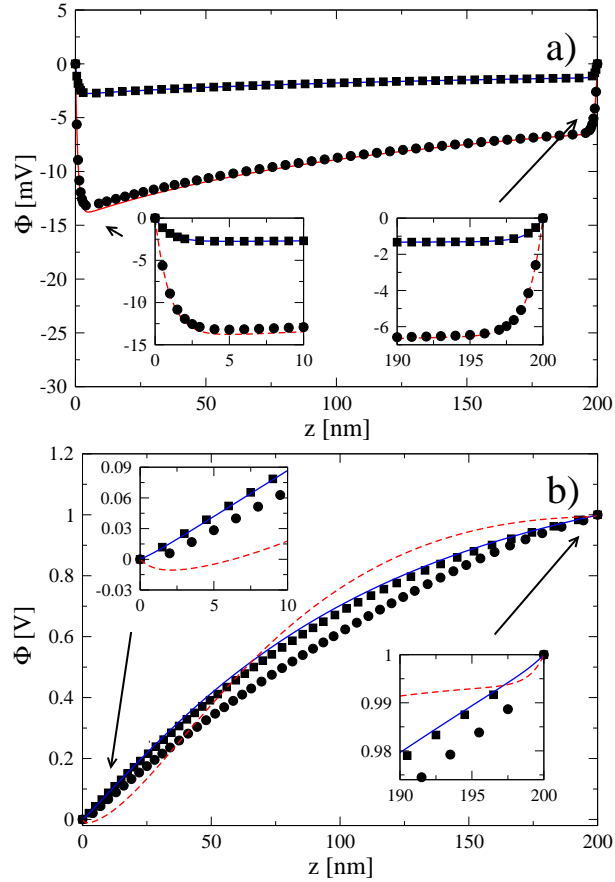

\begin{center}
 \includegraphics[width=8.0cm]{fig/pot_0.eps}
 \includegraphics[width=8.0cm]{fig/pot_1.eps}
\caption{(Color online) Potential profile $\Phi(z)$ for $c_{\mathrm{L}}=c_{\mathrm{R}}=0.1$ M. 
Calculations are done for $\Phi(0)=\Phi(L)=0$ V (a) and $\Phi(0)=0$ V, $\Phi(L)=1$ V (b).
Solid line and squares:
$\sigma=-0.02$ e/nm$^2$, dashed line and circles:
$\sigma=-0.1$ e/nm$^2$. Symbols in all cases stands for numerical 
solution, lines represent perturbation theory (first order in $\epsilon$). Insets depict 
closer look into left and right boundary layers.}
\label{pot_profile}
\end{center}
\end{figure}

\begin{figure}[htb!]
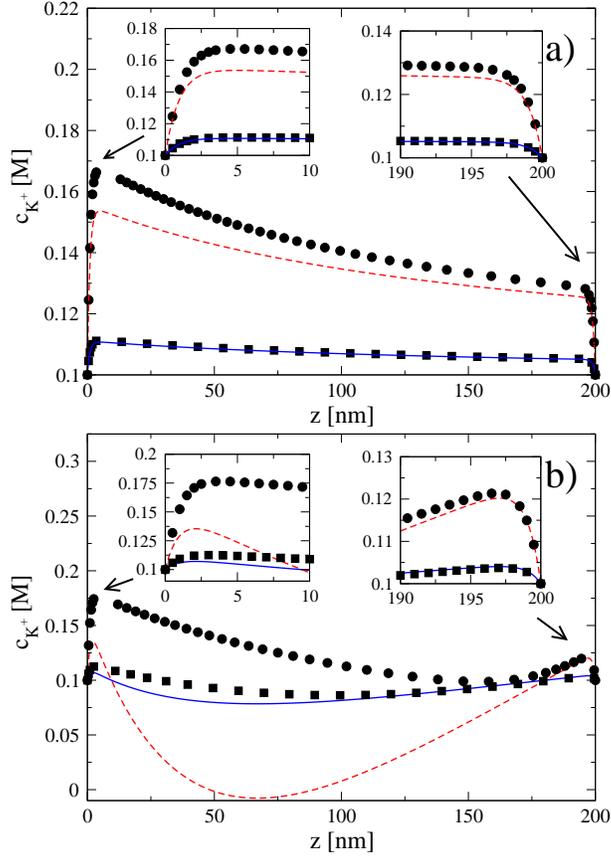

\begin{center}
 \includegraphics[width=8.0cm]{fig/con_pos_0.eps}
 \includegraphics[width=8.0cm]{fig/con_pos_1.eps}
\caption{(Color online) Concentration profile $c_{\rm{K}^+}(z)$ for $c_{\mathrm{L}}=c_{\mathrm{R}}=0.1$ M.
Calculations are done for $\Phi(0)=\Phi(L)=0$ V (a) and $\Phi(0)=0$ V, $\Phi(L)=1$ V (b).
Solid line and squares:
$\sigma=-0.02$ e/nm$^2$, dashed line and circles:
$\sigma=-0.1$ e/nm$^2$. Symbols in all cases stands for numerical 
solution, lines represent perturbation theory (first order in $\epsilon$). Insets depict 
closer look into left and right boundary layers.}
\label{con_pos_profile}
\end{center}
\end{figure}

\begin{figure}[htb!]
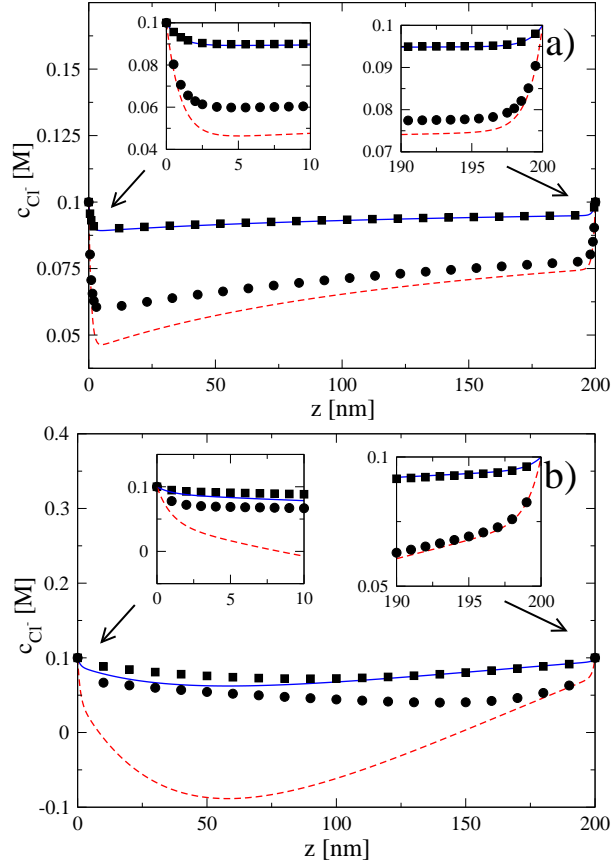

\begin{center}
 \includegraphics[width=8.0cm]{fig/con_neg_0.eps}
 \includegraphics[width=8.0cm]{fig/con_neg_1.eps}
\caption{(Color online) Concentration profile $c_{\rm{Cl}^-}(z)$ for $c_{\mathrm{L}}=c_{\mathrm{R}}=0.1$ M.
Calculations are done for $\Phi(0)=\Phi(L)=0$ V (a) and $\Phi(0)=0$ V, $\Phi(L)=1$ V (b).
Solid line and squares:
$\sigma=-0.02$ e/nm$^2$, dashed line and circles:
$\sigma=-0.1$ e/nm$^2$. Symbols in all cases stands for numerical 
solution, lines represent perturbation theory (first order in $\epsilon$). In sets depict 
closer look into left and right boundary layers.}
\label{con_neg_profile}
\end{center}
\end{figure}

The electric potential and concentration profiles for both the 
perturbation and the numerical solution are shown in Fig.~\ref{pot_profile}, Fig.~\ref{con_pos_profile}, 
and Fig.~\ref{con_neg_profile}. 
It is clearly shown that the agreement with the analytical result
(\ref{pot_uni}), and (\ref{con_pos_uni}) 
is getting worser with the increasing value of $\sigma$. 
Moreover, the discrepancies between theory and numerics are more
significant for growing absolute value of voltage. Figures~\ref{con_pos_profile}-\ref{con_neg_profile}
illustrate the situation in which the perturbation
approximation starts to miscarry for the charge density $\sigma=-0.1$
e/nm$^2$ in the non-equillibrium situation when the transmembrane
voltage reaches the value of 1 V. Note, in that case the
perturbation theory also provides unphysical negative
concentrations.

However, to assess the accuracy of power series in $\epsilon$, we shall systematically 
compare the approximations
including higher orders of this parameter with numerical solutions. 
Such a process for the uniformly valid solutions of singular perturbation theory 
(including boundary layer solutions)
is possible only up to second order in $\epsilon$. 
Nevertheless, on the basis of {\bf Theorem 2}, one can note that  
the outer expansions approximate well the solutions of the 1D-PNP problem 
with the Donnan boundary conditions.
One can obtain these perturbative expansions for any order in $\epsilon$. 

\subsubsection{Outer approximation}

\begin{figure}[htb!]
\begin{center}
 \includegraphics[width=8.0cm]{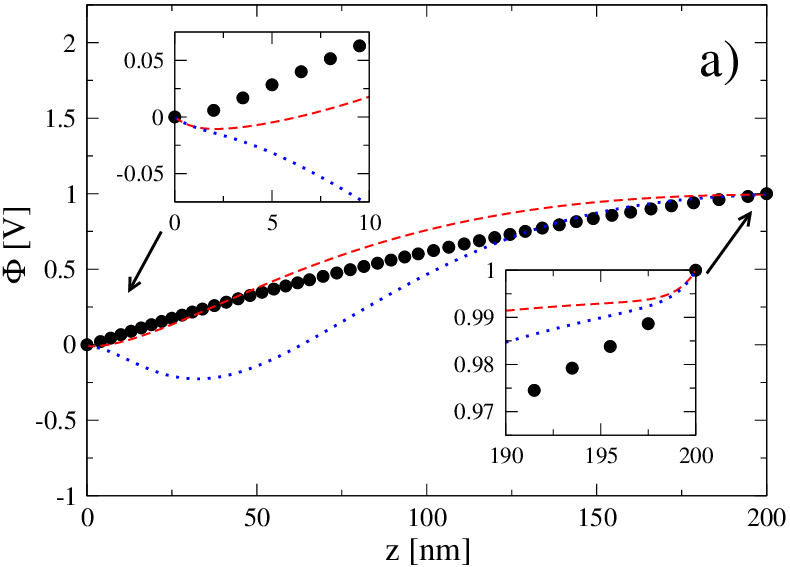}
 \includegraphics[width=8.0cm]{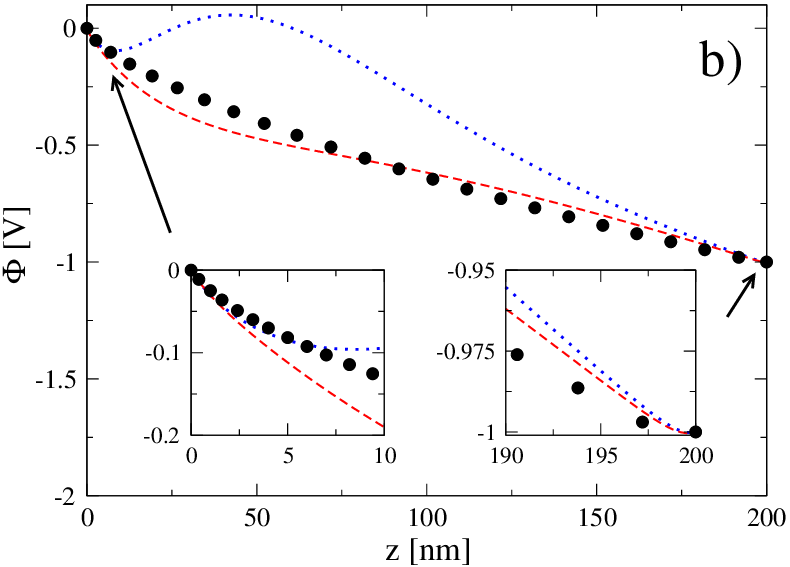}
\caption{(Color online) Potential profile $\Phi(z)$ for $c^{\rm{Don}}_{\rm{K}^+,~L} = 156$ mM, $c^{\rm{Don}}_{\rm{K}^+,~R} = 125$ mM, $c^{\rm{Don}}_{\rm{Cl}^-,~L} = 87.3$ mM, $c^{\rm{Don}}_{\rm{Cl}^-,~R} = 83.6$ mM, $\sigma=-0.1$ e/nm$^2$.
Calculations are done for $\Phi^{\rm{Don}}(0) = -0.014$ V, $\Phi^{\rm{Don}}(L) = 0.994$ V (a) 
and $\Phi^{\rm{Don}}(0)=-0.014$ V, $\Phi^{\rm{Don}}(L) = -1.006$ V (b).
We show perturbation solution in first order (in
   $\epsilon$): dashed line and second order:
   dotted line. The numerical solution is
   represented by symbols. Insets depict 
closer look into left and right boundary layers.}
\label{pot_dwa_profile}
\end{center}
\end{figure}

\begin{figure}[htb!]
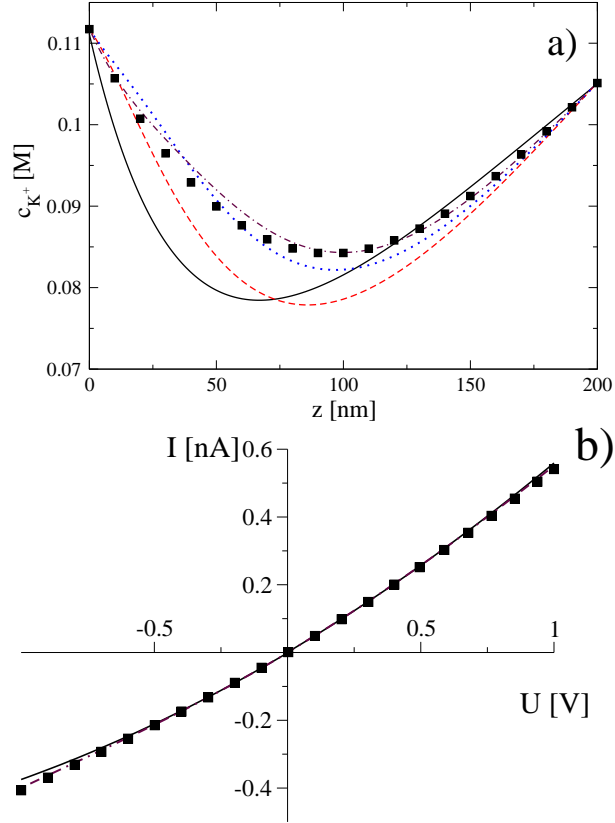

\begin{center}
 \includegraphics[width=8.0cm]{fig/con_pos_don_4.eps}
 \includegraphics[width=8.0cm]{fig/prad_k_4.eps}
\caption{(Color online) In the upper panel (a) we show the concentration profile 
$c_{K^+}(z)$ for $\sigma=-0.02$ e/nm$^2$, and the Donnan b. c.: $c^{\rm{Don}}_{\rm{K}^+,L} = 112$ mM, $c^{\rm{Don}}_{\rm{K}^+,R} = 105$ mM, $c^{\rm{Don}}_{\rm{Cl}^-,~L} = 131.8$ mM, $c^{\rm{Don}}_{\rm{Cl}^-,~R} = 103.8$ mM, $\Phi^{\rm{Don}}(0) = -0.003$ V, $\Phi^{\rm{Don}}(L)=0.999$ V;
and in the lower panel (b) we show the $I-U$ dependence.
We show perturbation solution in first order 
(in $\epsilon$): solid line, second order:
dashed line, third order:  dotted line, and fourth order: dash-dotted line. 
The numerical solution is represented by symbols.}
\label{pot_cztery_profile}
\end{center}
\end{figure}

In the non-equillibrium situation, 
we examine the electric potential profile for  $\epsilon=0.6$
(Fig.~\ref{pot_profile}b), and the concentration profile
for  $\epsilon=0.12$ (Fig.~\ref{con_pos_profile}b). 
Both approximations (in the first order of $\epsilon$) 
seem to show not too large discrepancy from the numerical solution.
However, if we plot the same expansion in Eq.~(\ref{pot_uni}) with
only one more term in the perturbation series, we notice a considerable deviation
from the numerical results one when the transmembrane voltage
becomes sufficiently 
large: $U = +1$ V, or $U = -1$ V (Fig.~\ref{pot_dwa_profile}).
This confirms that for the charge density $\sigma=-0.1$ the perturbation method
fails totally. However, for a smaller charge density $\sigma=-0.02$
the perturbation theory works as demonstrated  
in Fig.~\ref{pot_cztery_profile}, 
where the perturbation expansion was calculated up to the fourth order in 
$\epsilon$. 
One can notice that the agreement with numerics is getting better with 
growing order
of the perturbative expansion. Then,  
it yields very accurate approximations in all aspects: the potential 
and concentration profiles, 
the total flux, and the electric current as well.

\subsection{Rigorous b. c. vs. the Donnan b. c.}

\begin{figure}[htb!]
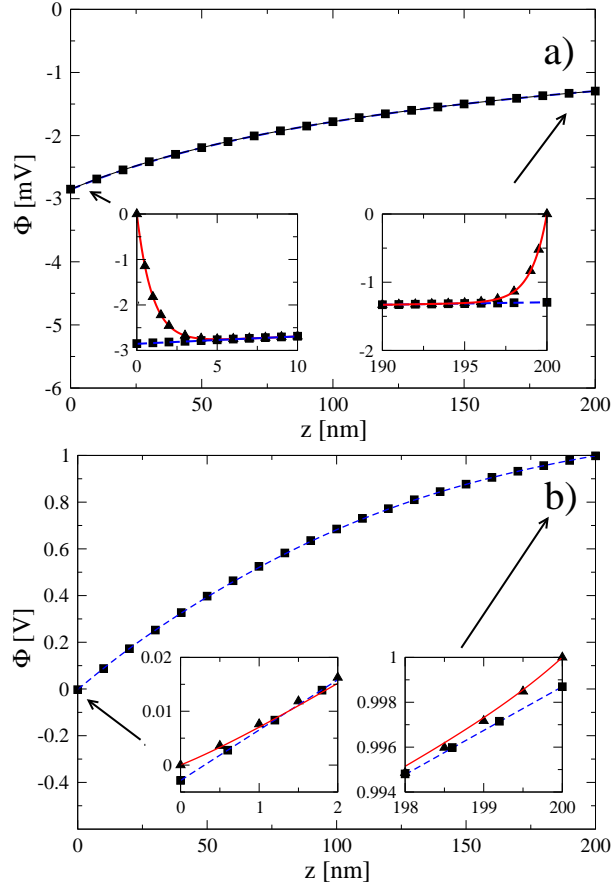

\begin{center}
 \includegraphics[width=8.0cm]{fig/pot_don_rig_0.eps}
 \includegraphics[width=8.0cm]{fig/pot_don_rig_1.eps}
\caption{(Color online) Potential profile $\Phi (z)$ for 
$c_L=c_R=0.1$ M, $\sigma=-0.02$ e/nm$^2$. 
Calculations are done for $\Phi(0)=\Phi(L)=0$ V (solid line)
and corresponding Donnan b. c. $\Phi^{\rm{Don}}(0) = -2.9$ mV, $\Phi^{\rm{Don}}(L) = -1.3$ mV (dashed line) (a);
$\Phi(0)=0$ V, $\Phi(L)=1$ V (solid line)
and corresponding Donnan b. c. $\Phi^{\rm{Don}}(0) = -0.003$ V, $\Phi^{\rm{Don}}(L) = 0.999$ V (dashed line) (b).
We show perturbation solution in first order 
(in $\epsilon$). Symbols in all cases stands for numerical 
solution, lines represent perturbation theory.}
\label{pot_don_rig}
\end{center}
\end{figure}

In Fig.~\ref{pot_don_rig}, we compare the rigorous boundary layer
solution with one given by the Donnan potential jump approximation.
The both agree well except from the behavior in the boundary layer, where
the "Donnan" solution makes naturally a jump
because it corresponds to the
approximation of the boundary layer of zero-width.   
Note the agreement with the {\bf Theorem 2}: the outer solutions agree very well.

\begin{figure}[htb!]
\begin{center}
 \includegraphics[width=8.0cm]{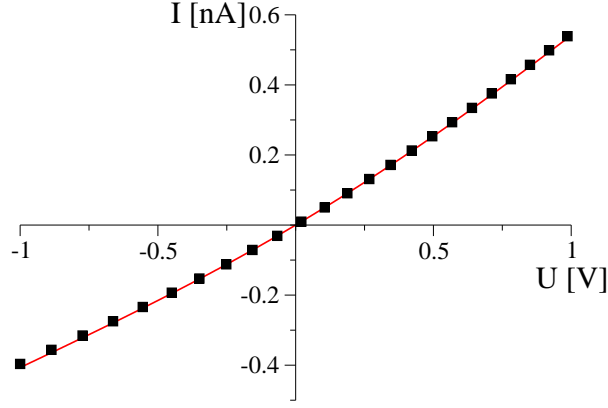}
\caption{
(Color online) Numerically obtained current voltage ($I-U$) characteristics 
for $\sigma=-0.02$ e/nm$^2$. Calculation are done for: rigorous b. c. (solid line) 
$c_{\mathrm{L}}=c_{\mathrm{R}}=0.1$ M and corresponding Donnan b. c. (squares) $c^{\rm{Don}}_{\rm{K^+,L}} = 112$ mM, $c^{\rm{Don}}_{\rm{K^+,R}} = 105$ mM, $c^{\rm{Don}}_{\rm{Cl^-,~L}} = 131.8$ mM, $c^{\rm{Don}}_{\rm{Cl^-,~R}} = 103.8$ mM.
}
\label{prad_don_rig}
\end{center}
\end{figure}

In Fig~\ref{prad_don_rig}, we also compare the numerical solutions for
the electric current in the case of the rigorous boundary conditions 
and in the case of the corresponding 
Donnan boundary condition approximation. The both solutions agree very well.

\section{Discussion}

The studied 1D PNP model describes ion flows in the conical geometry
in the presence of a permanent charge on the channel wall. The
dimensionality reduction from 3D to 1D results both in an   entropic
potential asymmetry and in the appearence of an inhomogeneous, asymmetric 
1D charge
density. In this paper, we presented the details  of the  singular
perturbation approach to the 1D PNP system including these profound effects. 
In another paper \cite{Kosinska}, these results are used further 
to find an analytical expression for
the electric current and to quantify the rectification effect. 
When the surface charge density is small enough, the
obtained theoretical expressions  are expected to provide good approximations
for the electric potential profile, concentrations of ions,  the total
flux, as well as for the electric current. 

Although such weakly charged channels seem to be less interesting 
from the experimental point of view, they are  important model systems
\cite{Kosinska}. In particular, based on our analytical solutions,
complemented also by the numerical analysis,  we studied and compared 
two different
boundary conditions: the rigorous boundary conditions and the
so-called  Donnan boundary condition.  The \emph{heuristic} assumption
of the Donnan equilibrium at the boundaries of the channel
\cite{Cervera}  in a highly non-equilibrium situation (current
rectification) should and must be questioned. However, our
analysis clearly  showed that when the channel length
exceeds much the Debye screening length, i.e. in the limit $1/\lambda
\to 0$, the use of the Donnan boundary conditions is well justified. 

\section*{Acknowledgments}

This work has been supported by
Volkswagen Foundation (project number I/80424),
the Alexander von Humboldt Foundation (I. D. K.),
the DFG (research center, SFB-486, project A10), and by the 
Nanosystems Initiative Munich.

\appendix

\section{1D model reduction}

Let us start from the 3D electro-diffusion equation
\begin{equation}
\displaystyle\frac{\partial c_i(\vec{r},t)}{\partial t} = -
\nabla \cdot \vec{j_i}(\vec{r},t)
\label{eq1}
\end{equation}
where
\begin{equation}\label{flux}
\vec{j_i}(\vec{r},t) = -D_i \nabla c_i(\vec{r},t)+
\rm{e}\nu_i \mu_i \vec{\mathcal{E}}(\vec{r}) c_i(\vec{r},t)
\end{equation}
defines the mass current $\vec{j_i}$ of the $i$-th species  of ions,
$\vec{r}$ is the position vector, $D_i$ is the diffusion
constant,  $\mu_i$ is the mobility of the ion particles, and
$\vec{\mathcal{E}}(\vec{r})=-\nabla \Phi(\vec{r})$ denotes the electric field. The consistency with
the thermal equilibrium demands that the mobility of the ions
$\mu_i$ fulfills the Einstein relation, reading
$D_i=\mu_i/\beta$, with the inverse temperature $\beta=1/k_{\rm{B}} T$.

The electric potential $\Phi({\vec{r}})$ is governed in a self-consistent manner by the Poisson
equation
\begin{equation}
\epsilon_0\nabla\cdot[\epsilon({\vec{ r}})\nabla\Phi({\vec{r}})] = -\sum_{i = \rm{K}^+, \rm{Cl}^-} \rho_i({\vec{ r}})  - \rho_{\rm{fix}}({\vec{r}}),
\label{Poisson}
\end{equation}
where $\rho_{\rm{fix}}({\vec{r}}) = \delta(r-R(z))\sigma$, with $\delta$ being the Dirac delta-function,
represents the fixed
charges located on the inside of the channel wall.
$\rho_i({\vec{ r}})$ denotes the density of mobile ions, and 
$\epsilon({\vec{ r}}) = \epsilon_p\Theta ( r - R(z) ) + \epsilon_w\Theta (-r + R(z))$ describes dielectric properties of the system, 
with $\epsilon_p,\epsilon_w$ being the relative dielectric constants of the polymer and water, respectively, and
$\Theta$ being the Heaviside step function. 
In above-presented definitions 
we made use of cylindrical coordinates that are natural for the geometry given in Fig.~\ref{schema}.

The set of relations (\ref{eq1}) and
(\ref{Poisson}) for $\vec{j_i}=\rm{const}$ constitutes the system of
coupled 3D Poisson-Nernst-Planck (PNP) equations. 

Under assumptions of instantaneous equillibration in the transverse direction 
as in Eq.~(\ref{constant}) and employing the definition of divergence given below 
(App. B) we derive the set of 1D PNP equations (Eqs.~(\ref{current}) and (\ref{poisson})), see also Ref.~\cite{Kosinska,Gardner}.

\section{The divergence theorem.}

The definition of the divergence of a vector $\vec{u}$ is
\begin{equation}
  \displaystyle\nabla\cdot \vec{u} = 
  \lim_{\mathcal{V}\to 0}\frac{1}{\mathcal{V}}\int_{\mathcal{S}}\vec{u}\cdot \hat{n}~dS
  = 
  \lim_{\Delta z\to 0}\frac{1}{A(z)\Delta z}\int_{\mathcal{S}} \vec{u}\cdot \hat{n}~dS,
\label{div}
\end{equation}
where $\mathcal{V}(z)=A(z)\Delta z$ denotes the volume surrounded by the
closed surface $\mathcal{S}$ with the normal vector $\hat{n}$. For the flux field
$\vec{j}(z) = j(z) \hat{e}_z$ (no ion flux across the channel
wall) Eq.~(\ref{div}) can be reduced \cite{Arfken} to the form
\begin{equation}
\displaystyle\lim_{\Delta z\to 0} 
\frac{1}{A(z)\Delta z}\left[ 
j(z+\Delta z)A(z+\Delta z) - j(z)A(z)] 
\right]
= \frac{1}{A(z)}\frac{d}{dz}(A(z)j(z)).
\end{equation}
In the case of the electric field we are intrested in the divergence: $\nabla\cdot\left(\epsilon(r,z)\vec{E}(z)\right)$.
When $ \vec{\mathcal{E}}(r,z) =  \mathcal{E}_z(r,z) \hat{e}_z + \mathcal{E}_r(r,z) \hat{e}_r$ the surface $S$ is either parallel or perpendicular
to $ \vec{\mathcal{E}}$. Inside the channel the electric field has only the $z$ component depending on $z$ coordinate: $ \vec{\mathcal{E}}(z) =  \mathcal{E}_z(z) \hat{e}_z$.
As the channel wall is charged with prescribed charge density $\sigma$, 
we find
\begin{equation}
\begin{array}{c}
\displaystyle
\vspace{10pt}
\lim_{\Delta z\to 0} 
\frac{\epsilon_w}{A(z)\Delta z}\bigg\{ 
\mathcal{E}(z+\Delta z)A(z+\Delta z) - \mathcal{E}(z)A(z)\bigg\}  \\
\vspace{10pt}
\displaystyle
- \left( \epsilon_{p} \mathcal{E}^{\perp}_{\rm{polymer}} 
- \epsilon_{w}\mathcal{E}^{\perp} \right)\frac{2}{R(z)} 
\displaystyle = \frac{\epsilon_w}{A(z)}\frac{d}{dz}(A(z)\mathcal{E}(z)) - 
\frac{2 \sigma}{\epsilon_0R(z)},
\end{array}
\end{equation}
where $A(z) = \pi R^2(z)$. The jump in the electric field is due to the surface charge density (the electrostatic boundary condition on the channel wall): 
\begin{equation}
\displaystyle \epsilon_p  \mathcal{E}^{\perp}_{\rm{polymer}} - \epsilon_w \mathcal{E}^{\perp} = \frac{\sigma}{\epsilon_0}.
\end{equation}
\section{The outer solutions - general expressions.}

\subsection{ Zeroth order in $\epsilon$}

In zeroth order in $\epsilon$, the solution reads
\begin{equation}
\displaystyle c^{(0)}_{(0),\mathrm{\Sigma}}(z)
 =
\left(\frac{-J^{(0)}_{(0)}}{R(z)\gamma} + C^{(0)}_{(0)}\right)R^{2}(z).
\label{con_0}
\end{equation}
Substituting that expression for the sum of concentrations in the second equation, we get
\begin{equation}
\displaystyle \Phi^{(0)}_{(0)}(z) = -I^{(0)}_{(0)}\int \frac{dz}{c^{(0)}_{(0), \mathrm{\Sigma}}(z)},
\label{pot}
\end{equation}
\begin{itemize}
\item
$J^{(0)}_{(0)} = 0$ if $c_\mathrm{L} = c_\mathrm{R}$ (see Appendix B)
\begin{equation}
\Phi^{(0)}_{(0)}(z) =\frac{ I_{(0)}^{(0)}}{C^{(0)}_{(0)}} \frac{1}{\gamma R(z)} + E^{(0)}_{(0)},
\end{equation}
\item
$J^{(0)}_{(0)} \ne 0$ if $c_\mathrm{L} \ne c_\mathrm{R}$
\begin{equation}
\Phi^{(0)}_{(0)}(z) =\frac{ I_{(0)}^{(0)}}{J^{(0)}_{(0)}}\ln \frac{R(z)}{-J^{(0)}_{(0)} + C^{(0)}_{(0)}\gamma R(z)} + E^{(0)}_{(0)}.
\end{equation}
\end{itemize}
\subsection{ First order in $\epsilon$}

The solution reads
\begin{itemize}
\item
$J^{(0)}_{(0)} = 0$
\begin{eqnarray}
 c^{(0)}_{(1), \mathrm{\Sigma}}(z) & = &
\bigg(
 -\frac{\pi I^{(0)}_{(0)}}{C^{(0)}_{(0)}}
\frac{1}{2\gamma R^{2}(z)}
- \frac{J^{(0)}_{(1)}}{\gamma R(z)}
 + C^{(0)}_{(1)}
\bigg) R^{2} (z)
\label{con_1_equal}
\end{eqnarray}
\item 
$J^{(0)}_{(0)} \ne 0$
\begin{equation}
\begin{array}{c}
\vspace{10pt}
\displaystyle 
c^{(0)}_{(1), \mathrm{\Sigma}}(z)=
\bigg\{\displaystyle \pi I^{(0)}_{(0)}\left( 
\frac{C^{(0)}_{(0)}\gamma}{J^{(0)~2}_{(0)}} 
\ln\left(\frac{-J^{(0)}_{(0)} + C^{(0)}_{(0)}\gamma R(z)}{ R(z)}\right) +
 \frac{1}{J^{(0)}_{(0)}R(z)} 
 \right)\\
\displaystyle - \frac{J^{(0)}_{(1)}}{\gamma R(z)}
 + C^{(0)}_{(1)} \bigg\} R^{2}(z).
\label{con_1}
\end{array}
\end{equation}
\end{itemize}
And the equation for the potential has the form
\begin{equation}
\displaystyle \Phi^{(0)}_{(1)}(z) = \int\frac{dz}{ c^{(0)}_{(0), \mathrm{\Sigma}}(z)}\left\{ -c^{(0)}_{(1), \mathrm{\Sigma}}(z)\frac{d\Phi^{(0)}_{(0)}(z)}{dz} - 
I^{(0)}_{(1)} + \pi R'(z) \right\}.
\label{pot_1}
\end{equation}

\subsection{ Higher orders in $\epsilon$}


The solution reads
\begin{equation}
\begin{array}{c}
\vspace{10pt}
\displaystyle 
c^{(0)}_{(2), \mathrm{\Sigma}}(z) =
\vspace{10pt}
\displaystyle \bigg(\mathcal{I}^{(0)}_{\{(0),(1)\}}(z)
- \frac{J^{(0)}_{(2)}}{\gamma R(z)}
 + C^{(0)}_{(2)} \bigg) R^{2}(z),
\label{con_2}
\end{array}
\end{equation}
where $\mathcal{I}^{(0)}_{\{(0),(1)\}}$ denotes the part of solution that does not depend on coefficients of second order in $\epsilon$
\begin{eqnarray}
\vspace{10pt}
\displaystyle \mathcal{I}^{(0)}_{\{(0),(1)\}}(z) & = & \int dz
\frac{c^{(0)}_{(1), \mathrm{\Sigma}}(z)}{R^2(z) c^{(0)}_{(0), \mathrm{\Sigma}}(z)}\left(  J^{(0)}_{(1)} -  \frac{d}{dz}c^{(0)}_{(1), \mathrm{\Sigma}}(z) + 2\frac{R'(z)}{R(z)}c^{(0)}_{(1), \mathrm{\Sigma}}(z)\right) \nonumber \\
& + & \int dz\frac{\pi\left(I^{(0)}_{(1)} - \pi R'(z)R(z)\right)}{R(z) c^{(0)}_{(0), \mathrm{\Sigma}}(z)}.
\label{i_dwa}
\end{eqnarray}

And the equation for the potential for $n\geq 2$ has the form
\begin{equation}
\displaystyle \Phi^{(0)}_{\mathrm{(n)}}(z) = \int\frac{dz}{ c^{(0)}_{(0), \mathrm{\Sigma}}(z)}\left\{ -\sum^{n-1}_{m=0}c^{(0)}_{\mathrm{(n-m)}, \mathrm{\Sigma}}(z)\frac{d\Phi^{(0)}_{\mathrm{(m)}}(z)}{dz} - I^{(0)}_{\mathrm{(n)}}\right\}.
\label{pot_n}
\end{equation}

The general solution for sum of concentrations for $n\geq 3$ reads
\begin{equation}
\begin{array}{c}
\vspace{10pt}
\displaystyle 
c^{(0)}_{\mathrm{(n)}, \mathrm{\Sigma}}(z) =
\displaystyle \bigg(\mathcal{I}^{(0)}_{\{(0),(1),\dots, \mathrm{(n-1)}\}}(z)
- \frac{J^{(0)}_{\mathrm{(n)}}}{\gamma R(z)}
 + C^{(0)}_{\mathrm{(n)}} \bigg) R^{2}(z),
\label{con_n}
\end{array}
\end{equation}
where 
\begin{equation}
\displaystyle \mathcal{I}^{(0)}_{\{(0),(1),\dots,\mathrm{(n-1)}\}}(z) = \sum^{n-1}_{m=1} \int dz
\frac{c^{(0)}_{\mathrm{(n-m)}, \mathrm{\Sigma}}(z)}{c^{(0)}_{(0), \mathrm{\Sigma}}(z) R^2(z)}\mathcal{B}_{\mathrm{m}} (z) + \int dz\frac{\pi I^{(0)}_{\mathrm{(n-1)}}}{ R(z) c^{(0)}_{(0), \mathrm{\Sigma}}(z)}
\label{I}
\end{equation}
with
\begin{equation}
\mathcal{B}_{\mathrm{m}} (z) = 
 J^{(0)}_{\mathrm{(m)}} -  \frac{d}{dz}c^{(0)}_{\mathrm{(m)}, \mathrm{\Sigma}}(z) + 2\frac{R'(z)}{R(z)}c^{(0)}_{\mathrm{(m)}, \mathrm{\Sigma}}(z).
\label{beta}
\end{equation}
The solution is iteratively given in orders of $\epsilon$.
Note that, the solution, which is valid in the interior of the channel (called the outer
expansion), is not valid near the both endpoints of the interval. Thus
we have to find a different representation for the solutions near
$z=0$ and $z=1$. These are boundary layer representation, which we
consider in the next section.  The outer expansion contain unknown
constants $J^{(0)}_{(i)}, I^{(0)}_{(i)},C^{(0)}_{(i)}, E^{(0)}_{(i)}$, where
$i=0,1,2,\dots$, which are determined by the matching condition.

\section{The outer solutions - determination of constants.}

We obtain analytical formulas for the constants. In zeroth order in $\epsilon$, in the cases when  
\begin{itemize}
\item $c_\mathrm{L} = c_\mathrm{R}=c$:
\begin{equation}
\begin{array}{cc}
\vspace{10pt}
\displaystyle J^{(0)}_{(0)} = 0,&
\displaystyle C^{(0)}_{(0)} = 2\pi c,\\
\vspace{10pt}
\displaystyle I^{(0)}_{(0)} = - C^{(0)}_{(0)}(1+\gamma)\Phi_\mathrm{R},&
\displaystyle E^{(0)}_{(0)} = \frac{(1+\gamma)}{\gamma}\Phi_\mathrm{R}.\\
\vspace{10pt}
\displaystyle J^{(0)}_{(1)} =
\frac{1}{2}(\gamma + 2)\pi\Phi_\mathrm{R},&
\displaystyle C^{(0)}_{(1)} = 
\frac{\pi}{2\gamma}\Phi_\mathrm{R},\\
\displaystyle I^{(0)}_{(1)} =
\frac{1}{12}\pi\gamma\Phi^2_\mathrm{R},&
\displaystyle E^{(0)}_{(1)} = 
- \frac{(\gamma + 2)}{24 c \gamma^2}\Phi^2_\mathrm{R}.
\end{array}
\end{equation}

\item $c_{\mathrm{L}} \ne c_\mathrm{R}$:
\begin{equation}
\begin{array}{cc}
\vspace{10pt}
\displaystyle J^{(0)}_{(0)} = 2\pi(1+\gamma)\left( c_{\mathrm{R}} - c_{\mathrm{L}} \right),&
\displaystyle C^{(0)}_{(0)} = \frac{2\pi}{\gamma}\left(
c_{\mathrm{R}}(1+\gamma)- c_{\mathrm{L}} \right),
\\
\vspace{10pt}
\displaystyle I^{(0)}_{(0)} = J^{(0)}_{(0)}
\frac{\Phi_\mathrm{R}}{\ln\left(c_{\mathrm{L}}/c_{\mathrm{R}}\right)},&
\displaystyle E^{(0)}_{(0)} = 
\frac{\ln(2\pi\gamma c_{\mathrm{L}})}{\ln(c_{\mathrm{L}}/c_{\mathrm{R}})}\Phi_\mathrm{R}.
\end{array}
\end{equation}
\end{itemize}
For higher orders these calculations get pretty quickly tedious.
Moreover, the formulas are very long.
Thus, we do not present them explicitly.
Instead, we propose a simple algorithm that we implemented in Maple.

Algorithm:

\begin{itemize}

\item

$c^{(0)}_{(0), \mathrm{\Sigma}}(z)$ (Eq.~(\ref{con_0})) and $\Phi^{(0)}_{(0)}(z)$ (Eq.~(\ref{pot}))

\item

$\displaystyle\mathcal{I}^{(0)}_{(0)}(z) = \pi I^{(0)}_{(0)} \int dz/\left(R(z)c^{(0)}_{(0), \mathrm{\Sigma}}(z)\right)$

\item

$\displaystyle J^{(0)}_{(1)} = (1+\gamma)\left( \mathcal{I}^{(0)}_{(0)}(0) - \mathcal{I}^{(0)}_{(0)}(1)  \right)$

\item

$\displaystyle C^{(0)}_{(1)} = J^{(0)}_{(1)}/\gamma - \mathcal{I}^{(0)}_{(0)}(0)$

\item

$\displaystyle c^{(0)}_{(1), \mathrm{\Sigma}}(z) = \left(\mathcal{I}^{(0)}_{(0)}(z) - J^{(0)}_{(1)}/\left(\gamma R(z)\right) + C^{(0)}_{(1)} \right) R^2(z)$

\item 

$\displaystyle \Phi^{(0)}_{(1)}(z)$ (Eq. (\ref{pot_1}))

\item

$\displaystyle I^{(0)}_{(1)} = \rm{solve} \left(\Phi^{(0)}_{(1)}(0) - \Phi^{(0)}_{(1)}(1) + \tilde{\phi}^{(0)}_{(1)}(-\infty) - \phi^{(0)}_{(1)}(\infty), I^{(0)}_{(1)} \right)$

\item

$\displaystyle E^{(0)}_{(1)} = \rm{solve} \left(\Phi^{(0)}_{(1)}(0) - \phi^{(0)}_{(1)}(\infty), E^{(0)}_{(1)} \right)$

\item

$\displaystyle\mathcal{I}^{ (0)}_{\{(0),\dots, \mathrm{n -1}\}}(z)$ 

- for $n=2$: Eq.~(\ref{i_dwa})

- for $n\geq 3$: Eqs.~(\ref{I}) and (\ref{beta})  

\item
\begin{eqnarray}
\vspace{10pt}
J^{(0)}_{\mathrm{(n)}} & = & (1+\gamma)
\bigg(
\displaystyle \mathcal{I}^{(0)}_{\{ 0,\dots, \mathrm{n - 1} \} } (0)- \mathcal{I}^{(0)}_{ \{ 0,\dots, \mathrm{n-1} \} }(1) \nonumber\\
\vspace{10pt}
&+& \displaystyle\frac{2}{(1+\gamma)^2}\sqrt{ \tilde{p}^{(0)}_{\mathrm{(n)}}(-\infty) \tilde{n}^{(0)}_{\mathrm{(n)}}(-\infty)} - 2\sqrt{ p^{(0)}_{\mathrm{(n)}}(\infty) n^{(0)}_{\mathrm{(n)}}(\infty)} \bigg) \nonumber
\end{eqnarray}

\item 

$\displaystyle C^{(0)}_{\mathrm{(n)}} = \frac{J^{(0)}_{\mathrm{(n)}}}{\gamma}  - \mathcal{I}^{(0)}_{\{0,\dots,\mathrm{n-1}\}} (0) + 2\sqrt{ p^{(0)}_{\mathrm{(n)}}(\infty) n^{(0)}_{\mathrm{(n)}}(\infty)}$.

\item

$c^{(0)}_{\mathrm{(n)}, \mathrm{\Sigma}}(z)$ (Eqs. (\ref{con_2}) and (\ref{con_n})) and $\displaystyle \Phi^{(0)}_{\mathrm{(n)}}(z)$ (Eq. (\ref{pot_n}))

\item

$\displaystyle I^{(0)}_{\mathrm{(n)}} = \rm{solve} \left(\Phi^{(0)}_{\mathrm{(n)}}(0) - \Phi^{(0)}_{\mathrm{(n)}}(1) + \tilde{\phi}^{(0)}_{\mathrm{(n)}}(-\infty)-\phi^{(0)}_{\mathrm{(n)}}(\infty), I^{(0)}_{\mathrm{(n)}} \right)$

\item

$\displaystyle E^{(0)}_{\mathrm{(n)}} = \rm{solve} \left(\Phi^{(0)}_{\mathrm{(n)}}(0) - \phi^{(0)}_{\mathrm{(n)}}(\infty), E^{(0)}_{\mathrm{(n)}} \right)$

\end{itemize}

In order to find values of $\bigg\{p^{(0)}_{\mathrm{(n)}}(\infty)$, $n^{(0)}_{\mathrm{(n)}}(\infty)$, $\phi^{(0)}_{\mathrm{(n)}}(\infty)\bigg\}$ and
$\bigg\{\tilde{p}^{(0)}_{\mathrm{(n)}}(-\infty)$, $\tilde{n}^{(0)}_{\mathrm{(n)}}(-\infty)$, $\tilde{\phi}^{(0)}_{\mathrm{(n)}}(-\infty)\bigg\}$ we make use of the series expansion of the Donnan equilibrium conditions (\ref{Donnan}) at $z_\mathrm{L}$ and $z_\mathrm{R}$, respectively (see {\bf Theorem 1} in Sec. 4).


\end{document}